%% file: main.tex
\documentclass[conference]{IEEEtran}
\IEEEoverridecommandlockouts
\usepackage{amsmath,amssymb,mathtools,amsthm,bm}
\usepackage{cite}
\usepackage{url}
\usepackage{physics}
\usepackage[hidelinks]{hyperref}
\usepackage{orcidlink}
\usepackage{enumitem}
\usepackage{booktabs}
\usepackage{comment}
\usepackage[caption=false,font=footnotesize]{subfig}

\usepackage{doi}
\usepackage{algorithm}

\usepackage{algpseudocode}
\usepackage{graphicx}
\usepackage{tikz}
\usetikzlibrary{shapes.geometric, arrows.meta, positioning, calc}
\usepackage{balance}
\usepackage[T1]{fontenc}

\usepackage{times}

\begin{document}
\addtolength{\textheight}{2\baselineskip}
\bstctlcite{IEEEexample:BSTcontrol}
\title{Large-Load Demand Flexibility as Virtual Storage \vspace{-0.25em}}

\author{%
\IEEEauthorblockN{%
Chandan~Chaudhary\orcidlink{0009-0002-2389-9568},~\emph{Student Member, IEEE},\\
Mohammed~Ben-Idris\orcidlink{0000-0002-8731-8913},~\emph{Senior Member, IEEE},
and Joydeep~Mitra\orcidlink{0000-0001-9287-0983},~\emph{Fellow, IEEE}%
}
\IEEEauthorblockA{%
Electrical and Computer Engineering, Michigan State University, East Lansing, MI 48824, USA\\
Emails: chaud152@msu.edu, benidris@msu.edu, and mitraj@msu.edu \vspace{-1.5em}}}

\maketitle

\begin{abstract}
\input{Contents/Abstract}
\end{abstract}


\IEEEpeerreviewmaketitle

\input{Contents/Introduction}
\input{Contents/Model}
\input{Contents/Method}
\input{Contents/Testsystem}
\input{Contents/conclusion}

\balance
\bibliographystyle{IEEEtran}
\bibliography{references}

\end{document}

%% file: Contents/Abstract.tex
Water electrolysis plants, hyperscale data centers, and aluminum potlines represent gigawatts of demand-side flexibility for bulk power system balancing, operational planning, and procurement services. Such loads are scheduled through per-interval power bounds and horizon energy windows, whereas co-located battery energy storage systems (BESS) operate under state-of-charge dynamics. The two formulations share no common mathematical structure, and the joint procurement value of co-located loads and storage goes unrealized as a result. This paper establishes the connection between the two formulations through a virtual storage (VS) equivalence. Every feasible large-load trajectory under power-bound and energy-window constraints is a valid charge trajectory of a VS device that operates at unity accounting efficiency in the grid power balance. Production and service-level costs lie outside this abstraction and enter the dispatch through curtailment opportunity costs. For a portfolio co-located with a BESS, aggregation reduces the constraint count from $O(NT)$ to $O(T)$, and its optimality conditions price the curtailment budget. Validation on the IEEE RTS-GMLC with three representative load classes shows that virtual storage delivers the dominant share of joint procurement savings. In the tested case, savings are additive because no constraint couples the two resources, and the curtailment shadow price equals the price at the marginal curtailment hour rather than the daily peak price.

%% file: Contents/Introduction.tex
\vspace{-0.5em}
\section{Introduction}
\label{sec:intro}
The growth of distributed energy resources (DER) and the changing role of dispatchable synchronous generation are reshaping the mix of resources used for balancing and ancillary services in bulk power systems~\cite{nerc2015essential, nerc2020fast}. Battery energy storage systems (BESS) and large flexible industrial loads have emerged as the two principal responses to the resulting adequacy and flexibility gap. However, these resources are scheduled through structurally different formulations. Large flexible loads are managed through per-interval power bounds and horizon energy windows within day-ahead market-clearing and resource-adequacy models, whereas BESS are operated through state-of-charge dynamics~\cite{ulbig2015analyzing}. The two formulations are incompatible and do not admit a unified dispatch, so significant system value goes unrealized. The joint operation of BESS and large flexible loads, particularly at the transmission level, remains an open problem.

Research on grid-scale storage has concentrated on optimizing charge and discharge decisions against uncertain DER output. On the demand side, flexibility models for residential, commercial, and industrial loads have been surveyed in~\cite{siano2014demand}. The aggregate flexibility of thermostatically controlled loads can be bounded by inner and outer generalized battery models with self-dissipation and two-way power limits~\cite{hao2014aggregate}, and virtual battery approximations of the Minkowski sum extend this result to heterogeneous populations~\cite{zhao2017geometric}. Flexibility envelopes have also been proposed as a planning tool that integrates supply-side and demand-side contributions~\cite{nosair2015flexibility}. However, none of these works establishes an exact equivalence between large-load curtailment trajectories and the charge trajectories of a charge-only storage device, nor do they provide a unified formulation to dispatch flexible loads jointly with co-located BESS.

Large industrial and commercial loads fall between these two strands of prior literature. Water electrolysis plants adjust power within seconds. A daily hydrogen production target sets their minimum energy floor~\cite{allidie2019pem, samani2020grid}. Hyperscale data centers can shed 5--10\% of rated load within 15 minutes. The actual curtailment depth depends on workload type and service-level constraints~\cite{wierman2014opportunities, liu2013data}. Aluminum potlines have provided balancing services for decades, and bidding strategies for energy and reserve auctions are well established~\cite{zhang2015bidding}. What these loads share, and what prior work has not formalized, is a curtailment cycle that structurally mirrors the charging phase of a BESS. At each interval, energy is withheld from the industrial process and the cumulative flexibility budget decreases monotonically with no energy-conversion loss.

Large-load growth has well-documented resource adequacy and dynamic consequences. Spatial correlation among large loads such as AI data centers inflates aggregate load variance and adequacy tail risk~\cite{chaudhary2026adequacy}, and it excites inter-area oscillations~\cite{Chaudhary2026SpatialAI, Chaudhary2026ModalSEST, Chaudhary2026CorrelatedDynamics}. The same workload coordination that correlates these loads also allows their curtailment to be scheduled as a single portfolio. Large loads with significant flexibility must therefore be dispatched jointly with co-located storage to extract their full grid value. The mathematical formulation to achieve this in a unified form does not yet exist.

The practical consequences of this structural gap take two forms. First, planning and scheduling models that co-locate large loads with BESS cannot form a unified linear program without a mathematical bridge between the two formulations~\cite{papavasiliou2014large}. Second, market mechanisms that clear load flexibility and storage through sequential, resource-specific processes assign independent shadow prices to each resource class. This sequential structure prevents joint intertemporal optimization and may create incentives for strategic capacity withholding by storage operators~\cite{rahimi2016transactive, kok2016society, ye2017strategic}. The joint formulation proposed in this paper removes this incentive in principle, though formal analysis of strategic behavior is left for future work.

This paper establishes the mathematical connection that enables coordinated operation of large flexible loads and co-located BESS in day-ahead scheduling and planning contexts. The main contributions are as follows.
\begin{enumerate}
\item The curtailment feasibility set of any large flexible load with a positive process floor, a rated power ceiling, and a horizon energy window is identical to the charge-trajectory set of a virtual storage (VS) device with parameters derived from the load's physical bounds. Unlike a BESS, VS is a charge-only construct.
\item For a portfolio of $N$ large loads co-located with a BESS, a Minkowski-sum aggregate representation reduces the load-side constraint count from $O(NT)$ to $O(T)$, independent of portfolio size. A single linear program jointly optimizes VS and BESS dispatch, and its optimality conditions price the curtailment budget at the net price of the marginal curtailment interval.
\item The method is evaluated on the IEEE RTS-GMLC~\cite{barrows2019ieee} with three large-load classes. Procurement savings, efficiency advantage, and constraint reduction are quantified relative to sequential dispatch.
\end{enumerate}

Section~\ref{sec:model} defines the large-load and BESS models, Section~\ref{sec:method} establishes the VS equivalence and co-dispatch formulation, Section~\ref{sec:results} presents the case study, and Section~\ref{sec:conc} concludes the paper.

%% file: Contents/Model.tex

\section{Large Load Flexibility and Modeling}
\label{sec:model}
Large flexible loads share a common scheduling structure in which power is bounded between a strictly positive process floor and a rated ceiling, with a horizon energy window as the sole intertemporal coupling. This section formalizes that structure.

\vspace{-0.25em}
\subsection{Physical Structure of Large Flexible Loads}
Three features distinguish large flexible loads from classical demand response. First, power must remain above a strictly positive process floor. The industrial or computational process cannot halt without physical damage or service failure. Second, total energy over the horizon must satisfy a minimum throughput requirement tied to the facility's production commitment. Third, consumption is freely adjustable within the admissible power range. These assumptions isolate the common structure needed for the VS mapping established in the methodology section.

\subsection{Power Bounds and Energy Window}
Let the planning horizon consist of $T$ intervals given by $\mathcal{T} = \{1, \ldots, T\}$, each of duration $\Delta t$ hours. The power consumption of a large flexible load at interval $t$ is $p^t$ in megawatts. At every interval, the load must operate between its process floor $\underline{p} > 0$ and its rated power $\bar{p}$:
\begin{equation}
\underline{p} \leq p^t \leq \bar{p}, \qquad \forall\, t \in \mathcal{T}.
\label{eq:pb}
\end{equation}
$\underline{p}$ is the minimum stable production load for an electrolyzer~\cite{allidie2019pem, matute2019technoeconomic, Niroula2023AlkalineElectrolyzer}, the thermal-balance floor for an aluminum potline~\cite{zhang2015bidding}, and the service-availability floor for a data center~\cite{wierman2014opportunities}.

Over the full horizon, the total energy consumed must satisfy the energy window constraint
\begin{equation}
E_{\min} \leq \sum_{t=1}^{T} p^t \,\Delta t \leq E_{\max},
\label{eq:ew}
\end{equation}
where $E_{\min}$ is the minimum throughput requirement and $E_{\max}$ is an upper consumption bound. The energy window is feasible if and only if
\begin{equation}
E_{\min} \leq T\bar{p}\,\Delta t, \qquad E_{\max} \geq T\underline{p}\,\Delta t, \qquad E_{\min} \leq E_{\max}.
\label{eq:feas}
\end{equation}
Throughout, $T\underline{p}\,\Delta t \leq E_{\min}$, so the throughput floor exceeds the minimum achievable consumption and the lower energy bound is nonredundant. $E_{\min}$ translates a production target into an energy equivalent and $E_{\max}$ is a contractual consumption ceiling~\cite{matute2019technoeconomic, liu2013data}. The load feasibility set is
\begin{equation}
\vspace{-0.75em}
\mathcal{F}_L = \bigl\{\,\mathbf{p} \in \mathbb{R}^T : \eqref{eq:pb} \text{ and } \eqref{eq:ew} \text{ hold}\,\bigr\}.
\label{eq:FL}
\end{equation}

\subsection{Deviation Variable and Curtailment Energy Bounds}
The operator-relevant quantity is the deviation from rated consumption. With the reference set at rated power $\bar{p}$ at every interval, the curtailment at interval $t$ is
\begin{equation}
\delta^t = \bar{p} - p^t \geq 0.
\label{eq:dev}
\end{equation}
In the nodal active-power balance, a verified load reduction of $\delta^t$~MW is equivalent to an injection of $\delta^t$~MW at the same bus. The flexibility depth $\Delta P = \bar{p} - \underline{p}$ gives $0 \leq \delta^t \leq \Delta P$ at every interval.

Substituting $p^t = \bar{p} - \delta^t$ into~\eqref{eq:ew} converts the energy window into a curtailment energy bound:
\begin{equation}
\vspace{-0.5em}
\underline{D} \leq \sum_{t=1}^{T} \delta^t \,\Delta t \leq \bar{D},
\label{eq:cd}
\end{equation}
where the curtailment energy bounds are
\begin{align}
\bar{D} &= T \bar{p} \Delta t - E_{\min}, \label{eq:Dbar}\\
\underline{D} &= \max\bigl\{0,\; T \bar{p} \Delta t - E_{\max}\bigr\}. \label{eq:Dund}
\end{align}
$\bar{D}$ is the maximum total curtailment without violating the throughput commitment. $\underline{D} \geq 0$ is the minimum curtailment when the load cannot operate at rated power for the full horizon. The deviation feasibility set is
\begin{equation}
\mathcal{F}_\delta = \bigl\{\,\boldsymbol{\delta} \in \mathbb{R}^T : 0 \leq \delta^t \leq \Delta P\ \forall t,\; \underline{D} \leq \mathbf{1}^\top\boldsymbol{\delta}\,\Delta t \leq \bar{D}\,\bigr\}.
\label{eq:Fd}
\end{equation}
From \eqref{eq:pb} and \eqref{eq:ew}, the mapping $\delta^t \! = \! \bar{p} - p^t$ is one-to-one between $\mathcal{F}_L$ and $\mathcal{F}_\delta$. Every feasible load trajectory corresponds to exactly one curtailment trajectory, and vice versa. The base model captures pure curtailment. The variable $\delta^t$ represents a reduction in consumption below rated draw, not a shift to later intervals. The lower bound $E_{\min}$ caps total curtailment at $\bar{D}$, so the production commitment holds without make-up consumption. For loads where the timing of recovery matters, the technology-specific extensions apply.

\vspace{-0.5em}
\subsection{Battery Energy Storage System}
A co-located BESS is characterized by charge power $u_c^t$, discharge power $u_d^t$, and state of charge (SOC) $e^t$~MWh at the end of interval $t$. The state of charge evolves as
\begin{equation}
e^{t} = e^{t-1} + \eta_c \Delta t\, u_c^t - \tfrac{\Delta t}{\eta_d}\, u_d^t,
\label{eq:soc}
\end{equation}
where $\eta_c, \eta_d \in (0,1]$ are the charge and discharge efficiencies, with $\eta_c \eta_d < 1$ assumed throughout. The physical limits are
\begin{align}
0 \leq u_c^t \leq \bar{u},\quad 0 \leq u_d^t \leq \bar{u}, &\quad \forall\,t \in \mathcal{T}, \label{eq:bpow}\\
\underline{e} \leq e^t \leq \bar{e}, \quad \forall\, t = 0,\ldots,T, &\quad e^0 = e^T = e_0, \label{eq:bsoc}
\end{align}
where $\bar{u}$ is the inverter power rating, $[\underline{e}, \bar{e}]$ is the usable SOC window, and $e_0$ is the initial and terminal state of charge.

\vspace{-0.2em}
\subsection{Technology-Specific Extensions}
\label{sec:ext}
The base model in~\eqref{eq:pb}--\eqref{eq:Fd} captures the dominant intertemporal coupling for large flexible loads. Three technology-specific constraints extend this base when fidelity requires it.

\textit{Electrolyzers} fit the base model well. Power adjusts within seconds, and the daily hydrogen production target maps directly to $E_{\min}$. The main extension is a hot-standby mode that consumes approximately 2\% of rated power with no hydrogen production. The energy window then applies only to production-mode intervals~\cite{matute2019technoeconomic}.

\textit{Data centers} fit the base model under delay-tolerant workloads and within cooling and service-level bounds. Load shedding of 5--10\% within 15~minutes is not universal. This flexibility depth depends on workload type, service-level agreements, and cooling headroom~\cite{wierman2014opportunities, liu2013data}. Coincident-peak and demand charges are captured through an interval power cap rather than through $E_{\max}$.

\textit{Aluminum potlines} provide an approximate fit. The rolling thermal-energy constraint, which requires sufficient energy over every successive $\tau_l$-hour sub-window, is stricter than the single-horizon window in~\eqref{eq:ew}~\cite{zhang2015bidding}. The base model is a conservative lower bound on potline flexibility. The case study uses the base model for the potline and treats the rolling constraint as a direction for future refinement.

%% file: Contents/Method.tex

\section{Virtual Storage Equivalence and Co-Dispatch}
\label{sec:method}

\subsection{The Virtual Storage Device}
Because $\delta^t \geq 0$ at every interval, the cumulative curtailment energy from the start of the horizon through interval $t$,
\vspace{-0.5em}
\begin{equation}
s^t = \sum_{\tau=1}^{t} \delta^\tau \,\Delta t,
\label{eq:vs}
\end{equation}
is nondecreasing. It starts at zero, rises as curtailments accumulate, and ends at the total curtailment energy $s^T \in [\underline{D}, \bar{D}]$. This is the trajectory of a charge-only accounting device. Curtailment draws down the load's finite flexibility budget monotonically, with no discharge. Because the sequence is nondecreasing, the terminal value $s^T \leq \bar{D}$ implies every intermediate value also satisfies $s^t \leq \bar{D}$, so intermediate capacity bounds are automatically satisfied. The \emph{virtual storage (VS)} device is defined by this charge-only trajectory, with initial state $s^0 = 0$, power rating $\Delta P$, capacity $C_\mathrm{VS} = \bar{D}$, minimum terminal energy $\underline{D}$, and unity accounting efficiency. Its feasibility set is
\begin{equation}
\mathcal{F}_\mathrm{VS} = \left\{(\boldsymbol{\delta}, \mathbf{s}) \;\middle|\;
\begin{array}{l}
s^t = \displaystyle\sum_{\tau=1}^{t} \delta^\tau \Delta t, \quad \forall\, t \in \mathcal{T},\\[4pt]
0 \leq \delta^t \leq \Delta P, \quad \forall\, t \in \mathcal{T},\\[4pt]
0 \leq s^t \leq \bar{D}, \quad \forall\, t \in \mathcal{T},\\[4pt]
\underline{D} \leq s^T \leq \bar{D}
\end{array}
\right\}.
\label{eq:FVS}
\end{equation}

\textit{Virtual Storage Equivalence:} In the base scheduling abstraction, the projection of $\mathcal{F}_\mathrm{VS}$ onto the deviation coordinates coincides exactly with $\mathcal{F}_\delta$. The set of all curtailment trajectories consistent with the base model is identical to the set of charge trajectories of the VS device.

The argument follows directly from the monotonicity observation above. Any $\boldsymbol{\delta} \in \mathcal{F}_\delta$ produces a nondecreasing $\mathbf{s}$ satisfying $s^t \leq s^T \leq \bar{D}$ and $s^T \in [\underline{D}, \bar{D}]$, so $(\boldsymbol{\delta}, \mathbf{s}) \in \mathcal{F}_\mathrm{VS}$. Conversely, any $(\boldsymbol{\delta}, \mathbf{s}) \in \mathcal{F}_\mathrm{VS}$ has $s^T = \mathbf{1}^\top \boldsymbol{\delta}\,\Delta t \in [\underline{D}, \bar{D}]$ and power bounds satisfied by definition, so $\boldsymbol{\delta} \in \mathcal{F}_\delta$. The equivalence is exact within the base model and requires no approximation.

The unity efficiency is an accounting convention of the base model, not a statement about physical recovery dynamics or economic cost. In the grid power balance, one megawatt of verified load reduction reduces net demand by exactly one megawatt, and no electrochemical conversion step is involved. Curtailment is not free, however. Production penalties, service-level violations, thermal recovery requirements, and rebound effects lie outside the base abstraction and enter through calibrated energy-window parameters, curtailment opportunity costs, or technology-specific constraints. A BESS cycles energy through charge and discharge with round-trip losses, so it can shift energy in time and arbitrage prices. VS is a charge-only budget. It cannot return energy to the grid or the load process, its capacity $\bar{D}$ is set by the production contract rather than by capital investment, and the budget resets at the start of each horizon.

\vspace{-0.75em}
\subsection{Aggregate Flexibility for a Portfolio of Loads}
For a portfolio of $N$ large flexible loads with individual deviation sets $\mathcal{F}_{\delta,i}$ parametrized by $(\Delta P_i, \underline{D}_i, \bar{D}_i)$, the aggregate flexibility is the Minkowski sum $\mathcal{F}_\mathrm{agg} = \bigoplus_{i=1}^N \mathcal{F}_{\delta,i}$. The additive aggregate parameters
\begin{equation}
\Delta P_\mathrm{agg} = \sum_i \Delta P_i, \quad \underline{D}_\mathrm{agg} = \sum_i \underline{D}_i, \quad \bar{D}_\mathrm{agg} = \sum_i \bar{D}_i
\label{eq:agg}
\end{equation}
define the outer set $\mathcal{F}_\mathrm{outer} = \{\boldsymbol{\delta} : 0 \leq \delta^t \leq \Delta P_\mathrm{agg},\, \underline{D}_\mathrm{agg} \leq \mathbf{1}^\top\boldsymbol{\delta}\,\Delta t \leq \bar{D}_\mathrm{agg}\}$.

\textit{Aggregation:} $\mathcal{F}_\mathrm{agg} \subseteq \mathcal{F}_\mathrm{outer}$. If the loads satisfy the proportionality condition $\underline{D}_i / (\Delta P_i \Delta t) = \alpha$ and $\bar{D}_i / (\Delta P_i \Delta t) = \beta$ for all $i$, then $\mathcal{F}_\mathrm{agg} = \mathcal{F}_\mathrm{outer}$. Under this sufficient condition, the proportional allocation $\delta_i^t = (\Delta P_i / \Delta P_\mathrm{agg})\,\delta_\mathrm{agg}^t$ is individually feasible for every load $i$.

Aggregation is exact when all loads share the same normalized flexibility profile. Each load allocates the same fraction of its flexibility to minimum throughput and maximum curtailment. When profiles differ, the outer set overstates what the portfolio can simultaneously deliver, and $\mathcal{F}_\mathrm{outer}$ is a strict outer approximation. Any aggregate dispatch must then pass a disaggregation feasibility check before individual curtailment targets are assigned. When disaggregation is infeasible, the operator may solve a second-stage disaggregation-constrained LP to find the nearest feasible individual schedule, or may restrict the aggregate bounds to the tighter inner approximation before scheduling.

\subsection{Joint Co-Dispatch Formulation}
With the aggregate VS representation in place, the joint scheduling problem for $N$ large loads and one co-located BESS over horizon $\mathcal{T}$ is formulated as a single linear program. Let $c^t$ denote the locational marginal price and $q^t \geq 0$ the curtailment opportunity cost at interval $t$. The total grid procurement is the rated load $\bar{p}_\mathrm{tot} = \sum_i \bar{p}_i$ reduced by aggregate curtailment and BESS net injection. The cost-minimization problem is
\begin{equation}
\min_{\boldsymbol{\delta}_\mathrm{agg},\,\mathbf{u}_c,\,\mathbf{u}_d,\,\mathbf{e}} \;\; \sum_{t=1}^T \Bigl[ c^t \bigl(\bar{p}_\mathrm{tot} - \delta^t_\mathrm{agg} - u_d^t + u_c^t\bigr) + q^t \delta^t_\mathrm{agg} \Bigr]\,\Delta t,
\label{eq:obj}
\end{equation}
subject to the aggregate load constraint $\boldsymbol{\delta}_\mathrm{agg} \in \mathcal{F}_\mathrm{outer}$, which by~\eqref{eq:agg} requires
\begin{equation}
0 \leq \delta^t_\mathrm{agg} \leq \Delta P_\mathrm{agg} \;\;\forall t, \qquad \underline{D}_\mathrm{agg} \leq \mathbf{1}^\top \boldsymbol{\delta}_\mathrm{agg}\,\Delta t \leq \bar{D}_\mathrm{agg},
\label{eq:lc}
\end{equation}
and the BESS constraints~\eqref{eq:soc}--\eqref{eq:bsoc}.

\textit{Constraint Reduction:} Each load $i$ contributes $2T$ interval power-bound and two energy-window constraints, giving $2NT + 2N = O(NT)$ load-side constraints. The aggregate outer set~\eqref{eq:lc} replaces these with $2T + 2 = O(T)$ constraints after the one-time summation in~\eqref{eq:agg}. This reduction is exact under the proportionality condition and otherwise yields an outer approximation that requires ex-post disaggregation validation. The BESS constraints~\eqref{eq:soc}--\eqref{eq:bsoc} are also $O(T)$, so the co-dispatch LP is $O(T)$ overall.

The formulation in~\eqref{eq:obj} is a joint optimization. The flexible loads and BESS share a single objective function, and all constraints are enforced simultaneously. This contrasts with sequential dispatch, in which load-only and BESS-only problems are each solved independently against fixed exogenous prices. Let $\lambda_{\bar{D}} \geq 0$ denote the dual variable of the upper energy bound in~\eqref{eq:lc}, which is the marginal value of the portfolio curtailment budget. Suppose this bound binds, the lower energy bound is slack, and some interval $t^\ast$ has $0 < \delta^{t^\ast}_\mathrm{agg} < \Delta P_\mathrm{agg}$. The KKT stationarity condition for $\delta^{t^\ast}_\mathrm{agg}$ then gives
\begin{equation}
\lambda_{\bar{D}} = c^{t^\ast} - q^{t^\ast}.
\label{eq:price}
\end{equation}
Intervals with $c^t - q^t > \lambda_{\bar{D}}$ are curtailed fully, and intervals with $c^t - q^t < \lambda_{\bar{D}}$ are not curtailed. No constraint in~\eqref{eq:obj} couples VS and BESS, so the BESS is valued by its own state-of-charge duals and both resources respond to the same LMP $c^t$. Curtailment could be compensated at $\lambda_{\bar{D}}$, analogous to a capacity settlement price. When disaggregation is infeasible, $\lambda_{\bar{D}}$ is the dual of the outer approximation. Load-specific settlement prices and full market mechanism design are left for future work.

%% file: Contents/Testsystem.tex

\vspace{-0.5em}
\section{Case Study}
\label{sec:results}
This section evaluates the VS co-dispatch method on the IEEE RTS-GMLC test system. Procurement cost savings, efficiency advantage, and constraint reduction are quantified across four scheduling scenarios over a 14-day summer study period. LMPs are treated as exogenous day-ahead price signals that do not respond to dispatch, consistent with the price-taking assumption.

\vspace{-0.65em}
\subsection{Test System and Load Placement}
The IEEE RTS-GMLC~\cite{barrows2019ieee} is a 73-bus, three-area synthetic transmission system with 5{,}224~MW of wind and solar capacity. Day-ahead locational marginal prices are drawn from the PLEXOS allTX production-cost solution distributed with the dataset. 


A PEM electrolyzer rated at 200~MW is placed at bus~118 (Area~1), a hyperscale data center rated at 250~MW at bus~218 (Area~2), and an aluminum potline rated at 300~MW at bus~318 (Area~3), each at its area's highest-load bus (333~MW base peak). 
This placement co-locates each load with adequate transmission capacity. The inter-area portfolio fails the proportionality condition by design, which activates the outer-approximation disaggregation path described in the methodology section. A BESS rated at 200~MW / 400~MWh is placed at bus~118. This single-site configuration is representative of co-located large-load and storage deployments. The BESS parameters are $\eta_c = \eta_d = 0.95$ ($\eta_{rt} = 0.9025$) and initial and terminal SOC $e_0 = 200$~MWh. Extending to one BESS per load bus is a direction for future work. The scheduling horizon is $T = 24$ hourly intervals with opportunity cost $q^t = 0$. Table~\ref{tab:vs_params} gives the virtual storage parameters derived via~\eqref{eq:Dbar}--\eqref{eq:Dund}. Electrolyzer parameters follow~\cite{allidie2019pem, matute2019technoeconomic, Niroula2023AlkalineElectrolyzer}. Data-center flexibility depth follows~\cite{wierman2014opportunities, liu2013data}. Potline parameters follow~\cite{zhang2015bidding}. The numerical values represent design-point choices within ranges reported in these sources.
All LPs are solved with HiGHS~v1.7~\cite{huangfu2018parallelizing} through SciPy~1.13.

\begin{table}[!htbp]
\centering
\vspace{-1.5em}
\caption{VS Parameters (24-Hour Horizon)}
\label{tab:vs_params}
\vspace{-1em}
\begin{tabular}{lrrrrr}
\toprule
Load              & Bus & $\bar{p}$ & $\Delta P$ & $\underline{D}$ & $\bar{D}$   \\
                  &     & (MW)       & (MW)        & (MWh)            & (MWh)        \\
\midrule
PEM Electrolyzer  & 118 & 200        & 180         & 240              & 1{,}200      \\
Data Center       & 218 & 250        & 150         & 600              & 1{,}800      \\
Aluminum Potline  & 318 & 300        &  45         & 360              &   720        \\
\midrule
\textbf{Aggregate} & --- & \textbf{750} & \textbf{375} & \textbf{1{,}200} & \textbf{3{,}720} \\
\bottomrule
\end{tabular}
\end{table}

\begin{figure*}[!t]
\centering
\subfloat[Load-Only (VS only)]{%
    \includegraphics[width=0.325\textwidth]{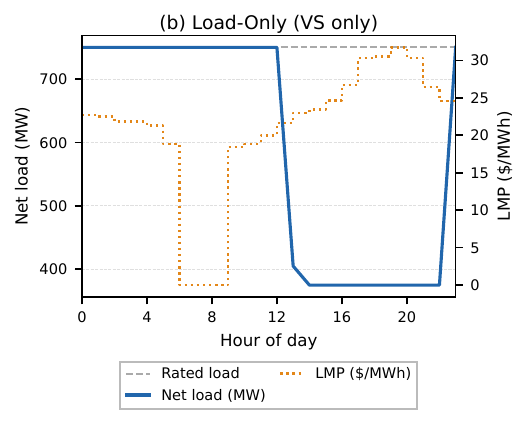}%
    \label{fig:disp_lo}}
\hfill
\subfloat[BESS-Only]{%
    \includegraphics[width=0.325\textwidth]{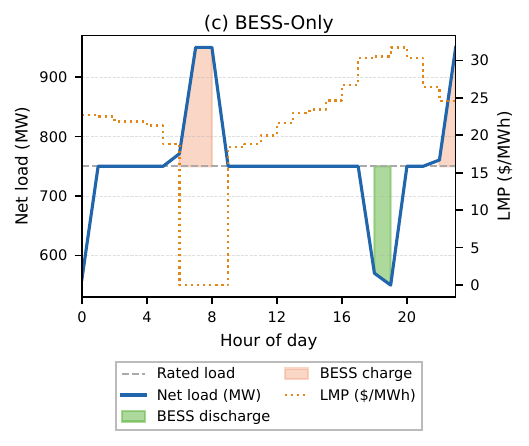}%
    \label{fig:disp_bo}}
\hfill
\subfloat[Co-Dispatch (VS + BESS)]{%
    \includegraphics[width=0.325\textwidth]{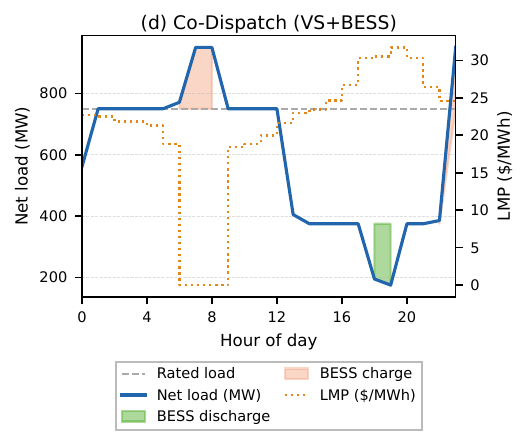}%
    \label{fig:disp_cd}}
\caption{Net load (MW) and LMP (\$/MWh) for three scheduling scenarios, July~5.
BESS charge/discharge intervals are shaded in (b) and~(c).}
\vspace{-1em}
\label{fig:dispatch_4scenarios}
\end{figure*}

\vspace{-0.5em}
\subsection{LMP Structure and Study Period}
The portfolio LMP is $c^t = (200\lambda_{118}^t + 250\lambda_{218}^t + 300\lambda_{318}^t)/750$. The 14-day study window (July~5--18, 2020) covers 336 hours with prices ranging from \$0 to \$111.59/MWh. Approximately 15\% of hours carry zero LMPs due to midday solar oversupply. Peak prices exceed \$80/MWh on five days in the window. The reference day for the single-day analysis is July~5, a standard summer day with a three-hour zero-price solar window (hours~6--8) and an evening peak of \$31.73/MWh at hour~19. This day is representative of the price structure that drives VS value throughout the study window.

\subsection{Dispatch Analysis: July 5 Reference Day}
Table~\ref{tab:results_july5} reports costs and procurement cost savings for the four scenarios. All percentage savings use the baseline procurement cost, with every load at rated power and no BESS, as the denominator. The load-only and BESS-only scenarios constitute sequential dispatch. Each resource is optimized independently with fixed exogenous prices. The co-dispatch scenario jointly optimizes both resources in a single LP. The load-only and co-dispatch scenarios operate on the aggregate VS outer representation. Individual load setpoints are not computed because the disaggregation LP is infeasible for the selected parameters. Fig.~\ref{fig:dispatch_4scenarios} shows the hourly net load trajectories overlaid on the LMP. Load-Only concentrates all 3{,}720~MWh of curtailment over the ten peak-price hours (13--22) and reduces net load by up to 375~MW during the evening ramp. BESS-Only charges at zero-price hours~6--8 and discharges 180--200~MW at the peak. Co-Dispatch superimposes both resources. Curtailment and BESS discharge coincide at hours~18--19 to achieve the maximum procurement reduction.

\begin{table}[!htbp]
\centering
\vspace{-1em}
\caption{Scheduling Results, July 5, 2020 (Baseline: Rated Load, No BESS)}
\label{tab:results_july5}
\vspace{-1em}
\begin{tabular}{lrrr}
\toprule
Scenario        & Daily Cost (\$)  & Savings (\$)  & Savings (\%) \\
\midrule
Baseline        & 376{,}650        & ---           & ---          \\
Load-Only (VS)  & 275{,}473        & 101{,}177     & 26.86        \\
BESS-Only       & 365{,}252        &  11{,}398     &  3.03        \\
Co-Dispatch     & 264{,}075        & 112{,}575     & 29.89        \\
\bottomrule
\end{tabular}
\vspace{-0.5em}
\end{table}


In the tested case, VS and BESS procurement cost savings are additive, as shown in Table~\ref{tab:results_july5}. This additivity holds because no constraint in the LP couples VS and BESS, so each resource follows the LMP independently even at hours~18--19, where curtailment and BESS discharge coincide. A shared interconnection limit or network constraints would couple the two resources, and additivity is not guaranteed in general.

Three quantitative findings follow. First, flexible loads are the dominant value source under zero opportunity cost. Second, the VS efficiency advantage is 362.7~MWh/day, which equals the $(1 - \eta_{rt})$ round-trip loss on the full 3{,}720~MWh curtailment budget. Third, portfolio aggregation reduces the load-side block from 150 ($O(NT)$) to 50 ($O(T)$) constraints, a factor of $N = 3$. Because $q^t = 0$, the reported savings include no industrial production penalty. Of the \$101{,}177 load-only savings, \$77{,}841 is the value of 3{,}720~MWh not purchased at the daily mean LMP, and \$23{,}336 comes from curtailing in high-price hours. Forgone production would offset part of the first component, so a nonzero opportunity cost would narrow the VS advantage.

\subsection{14-Day Study Summary}
\begin{figure}[!htbp]
\centering
\vspace{-1em}
\includegraphics[width=\columnwidth]{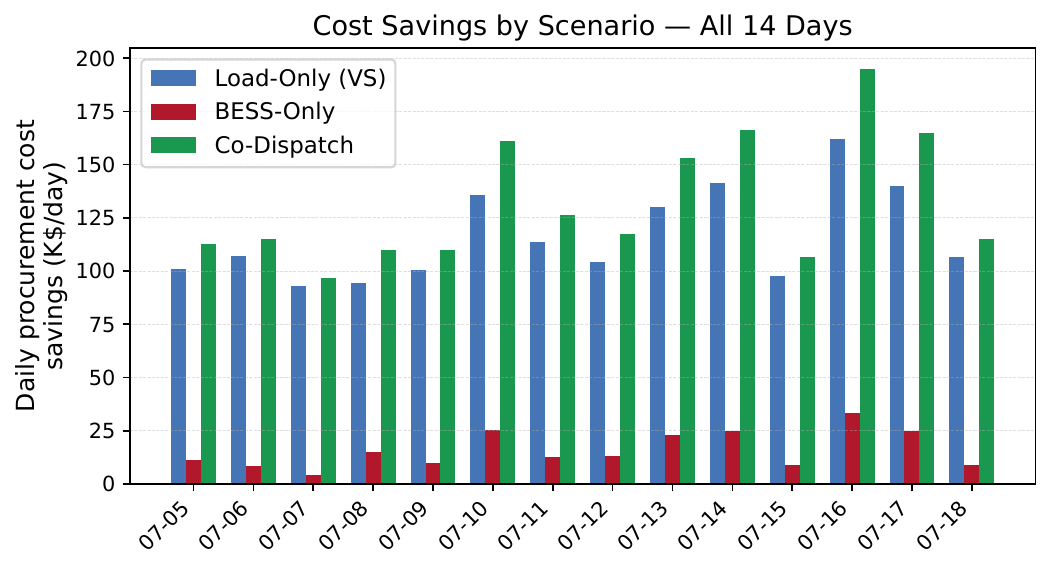}
\vspace{-2em}
\caption{Daily procurement cost savings by scenario, July~5--18.}
\vspace{-1em}
\label{fig:savings_14days}
\end{figure}

Fig.~\ref{fig:savings_14days} and Table~\ref{tab:14day_summary} report daily and aggregate procurement cost savings. Co-dispatch procurement cost savings total \$1{,}850{,}094 over 14 days against a baseline of \$5{,}475{,}478, a mean daily reduction of 33.9\%. On every day the LP exhausts the full 3{,}720~MWh curtailment budget, and the shadow price $\lambda_{\bar{D}}$ remains in the range \$21--\$26/MWh regardless of whether the daily peak price is \$30 or \$112/MWh. This stability follows from~\eqref{eq:price}. With $q^t = 0$, $\lambda_{\bar{D}}$ equals the LMP of the marginal curtailment hour, which is the tenth-highest hourly LMP of the day rather than the daily peak price. This range applies to the selected 14-day summer window. Stability across other seasons, load mixes, and price distributions remains to be studied. Disaggregation is infeasible on all 14 days, so the outer approximation is binding across the full study window. The conventional per-load formulation, which enforces the power and energy bounds of each load individually through 150 load-side constraints, yields a feasible schedule with 14-day load-only savings of \$1{,}604{,}871. The aggregate VS result of \$1{,}626{,}691 overstates this feasible value by 1.34\%.

\begin{table}[!htbp]
\centering
\vspace{-1em}
\caption{14-Day Aggregate Results (July 5--18, 2020)}
\label{tab:14day_summary}
\resizebox{\columnwidth}{!}{%
\vspace{-1em}
\begin{tabular}{lrrr}
\toprule
Metric                              & Load-Only       & BESS-Only       & Co-Dispatch     \\
\midrule
Total savings (\$)                  & 1{,}626{,}691   & 223{,}403       & 1{,}850{,}094   \\
Mean daily savings (\%)             & 29.84           & 4.06            & 33.90           \\
Min / Max daily savings (\%)        & 23.24 / 38.30   & 1.01 / 7.85     & 24.24 / 46.16   \\
Efficiency advantage (MWh/day)      & --              & --              & 362.7           \\
$\lambda_{\bar{D}}$ range (\$/MWh)  & --              & --              & 21.12--26.32    \\
Disaggregation feasible             & --              & N/A             & Never (14/14)   \\
\bottomrule
\end{tabular}%
}
\vspace{-1em}
\end{table}

\subsection{Discussion}
Three structural implications follow from the case study.
\textit{The virtual storage budget is set by industrial production economics, not by grid investment.} BESS capacity is a capital decision priced in \$/kWh. A potline's or electrolyzer's flexibility budget is set by its production contract. The load must consume a minimum energy over the horizon regardless, so the room between the throughput floor and rated draw is flexibility the industrial process has already paid for. This is why the aggregate VS budget exceeds the 400~MWh BESS by nearly an order of magnitude.

\textit{The shadow price of the curtailment budget is a candidate settlement signal.} $\lambda_{\bar{D}}$ equals the LMP at the marginal curtailment hour, not the daily peak price. Because the VS exhausts its budget on every study day, $\lambda_{\bar{D}}$ reflects the systematic afternoon peak-band structure. A curtailment contract priced at $\lambda_{\bar{D}}$ would be self-triggering on any standard summer day without a spike forecast. This differs from a peak-price-indexed signal, which transfers spike-day risk to the load operator. Formal settlement mechanism design and broader empirical validation are outside the scope of this paper.

\textit{Disaggregation infeasibility is diagnostic information about portfolio composition.} The outer approximation $\mathcal{F}_\mathrm{outer}$ strictly contains $\mathcal{F}_\mathrm{agg}$ because the three loads have dissimilar normalized flexibility. A portfolio with matched normalized energy windows would satisfy the proportionality condition and make the aggregate representation exact. Electrolyzer and potline energy windows are set by production contracts, so an operator can negotiate them toward proportionality.

Although the exogenous nodal LMPs reflect marginal congestion conditions from the underlying market solution, the present co-dispatch model does not feed VS and BESS decisions back into network flows. The reported savings therefore represent price-taking economic value rather than a network-constrained feasibility result. A network-constrained extension will embed the co-dispatch LP within a DC optimal power flow.

%% file: Contents/conclusion.tex

\section{Conclusion}
\label{sec:conc}
This paper has established that the curtailment feasibility set of any large flexible load under power-bound and energy-window constraints is identical to the charge-trajectory set of a VS device. This equivalence reduces joint large-load and BESS co-dispatch to a single LP with load-side constraints that scale with the horizon length independently of portfolio size. On the IEEE RTS-GMLC, VS delivered the larger share of procurement savings because its curtailment budget, which derives from existing production commitments, exceeds the BESS energy capacity by nearly an order of magnitude. The curtailment shadow price remained stable because it equals the LMP at the marginal curtailment hour, which makes it a candidate signal for day-ahead curtailment settlement. These results assumed zero curtailment opportunity cost and exogenous LMPs, so the savings quantify price-taking procurement value rather than network-constrained dispatch value.
